\documentclass[11pt]{amsart}
\usepackage{amsmath,amsfonts,amssymb}
\usepackage{microtype}
\usepackage{graphicx}
\usepackage{color}
        \definecolor{orange}{RGB}{255,127,0}
\usepackage{amsaddr}
\usepackage{hyperref}
\usepackage[alphabetic,backrefs]{amsrefs}
\begin{document}

\title{Non-Euclidean ideal spectrometer}

\author{Henrique N. S\'a Earp and Vladmir Sicca}
\address{IMECC - Institute of Mathematics, 
Unicamp. }
\author{Bernardo B. C. Kyotoku}
\address{IFGW - Institute of Physics, Unicamp.}
\email{henrique.saearp@ime.unicamp.br, ra083067@ime.unicamp.br}
\email{kyotoku@ifi.unicamp.br}
\date{\today}
\maketitle

\begin{abstract}
We describe the mathematical scheme for an anomaly-free ideal spectrometer, based  on a $2-$dimensional plane medium with conical regions of bounded slope. Moreover, the construction may be realised in many different configurations. \end{abstract}


\tableofcontents


\section{Introduction}

Advances in non-Euclidean optics over last decade include the development of several new optical devices, as well as the drastic improvement of existing ones \cite{pendry_controlling_2006,leonhardt_optical_2006,li_hiding_2008}.  However, spectrometers have not hitherto been significantly examined  in this perspective, besides a few interesting  proposals of alternative geometries (see \cite{ huang-aberration-2008, shi_three-focal-point_2002,  rowland_concave_1883,  horst_silicon--insulator_2009}). Crucially, it is known that such components always present aberrations under Euclidean geometry \cite{beutler_theory_1945,marz_theory_1992,namioka_theory_1959}. For a more detailed account of this difficulty, we refer the reader to the second named author's thesis \cite{Kyothesis}.
In this paper we show how a certain non-Euclidean optical medium solves a geometric problem of path-length constraints formulated in \cite{Kyothesis}. It can be interpreted as a first approximation to a mathematically  aberration-free spectrometer design, which may hopefully be exploited by optical scientists. Moreover, our construction has many degrees of freedom in the actual array of samples and groove points. 
The technique we use is the grafting of locally conical geometry onto an a priori flat optical medium. This idea stems from two interesting papers in the theoretical context of smectics on multi-layered media in mathematical relativity \cite{Mosna_2012,Mosna_2012b}. We thank R. Mosna for bringing these references to our attention.  

\subsection{Statement of the problem}

Given $m,n\in\mathbb{N}$ and a discrete set of positive real numbers $\Lambda:=\left\{0<\lambda_{1}<...<\lambda_{m}\right\}\subset\mathbb{R}$,  an \emph{(ideal) spectrometer of range $\Lambda$ and $n$ grooves} is defined as an almost everywhere smooth surface $\sigma\subset \mathbb{R}^3$ and a collection of labeled points $R_0,R_{1},...,R_{m},G_{1},...,G_{n}\in\sigma $  with the following physical significance:  
$$
 \begin{array}{cl}         \sigma & \text{optical medium}\\         \lambda_{1},...,\lambda_{m} & \text{wavelengths}\\         O:=R_{0}& \text{source of light}\\         R_{1},...,R_{m} & \text{receptors}\\         G_{1},...,G_{n} & \text{grating grooves.}\\ \end{array} 
$$
 Notice that each wavelength $\lambda_i$ is detected by a specific receptor $R_i$.
We search an array of the $\left\{G_{k}\right\}$ and $\left\{R_{i}\right\}$ such that a different beam of light emitted at $O$ crosses \emph{each} sample and reaches \emph{every} receptor at a  \emph{distinct phase} prescribed by constants $b_{k,i}\in\mathbb{N}$ (assumed ordered in the sense that $p>k\Rightarrow b_{p,i}>b_{k,i}$):
\begin{equation}        \label{eq: constraint}         \overline{OG_{k}}+\overline{G_{k}R_{i}}=b_{k,i}\lambda_{i},         \quad\forall i\in{\{1,...,m\}} \end{equation} where $\overline{AB}$ denotes the geodesic length of segment $AB$ over the surface $\sigma$.

\subsection{Structure of the paper} \label{procedimento}

We organise the construction problem in two steps. 
First, we study possible arrays for a spectrometer on a general $2-$dimensional optical medium $\sigma$ with uniform refractive index and show that the degrees of freedom by far outnumber the constraints.  We obtain a locally Euclidean solution with small conical regions of geodesic reflection. As a practical application, we propose a detailed example of a neat modular linear array.
Second, the design is conformally mapped to a plane optical medium with a non-uniform refractive index, following the complex-analytic methods of \cite{leonhardt_optical_2006,li_hiding_2008}. We exploit the mathematical fact that such material media of varying refraction  emulate precisely the differential-geometric effects of surface curvature. 
\section{General solution}
\label{sec general solution}
We now describe the general procedure to devise ideal spectrometer arrays. Consider first an array of receptors $\left\{R_{i}\right\}$ on a plane medium $\sigma_0\subset\mathbb{R}^2$ interspaced by Euclidean distances  
$$
 d_{i}:=\overline{R_{i-1}R_{i}}. 
$$
 The question is how to rearrange the $R_i$ so as to be able to place the groove points $G_k$ according to the prescribed constraint (\ref{eq: constraint}), changing the local geometry of $\sigma_0$ if necessary. 

\subsection{Elliptic reflection}
\label{Elipse}

Since the constants $b_{k,i}$ and the wavelengths $\lambda_i$ are strictly increasing (in $i$), upon assigning a sufficiently small real value $\epsilon_{1,i}>0$  to each  $b_{1,i}$, we may reposition the receptors $R_i$ closer to each other in such a way that, for each $i$, 
$$
 d_i<b_{1,i}\lambda_{i}-b_{1,i-1}\lambda_{i-1}-\epsilon_{1,i}. 
$$
 Moreover, we may choose small enough $\epsilon_{k,i}>0$ and define a family of ellipses $E_{k,i}$ of foci $R_{i-1}$ and $R_{i}$ for each $i\in\left\{ 1,...,m\right\} $ as the sets 
$$
 E_{k,i}:=\left\{ P\in\mathbb{R}^{2} \;|\; \overline{R_{i-1}P}+\overline{PR_{i}}=a_{k,i}\right\} 
$$
 with $a_{k,i}:=b_{k,i}\lambda_{i}-b_{k,i-1}\lambda_{i-1}-\epsilon_{k,i}$.

\begin{figure}[!ht] 
\label{fig: ellipses and example} 
\center{ 
\includegraphics[height=5.7cm]{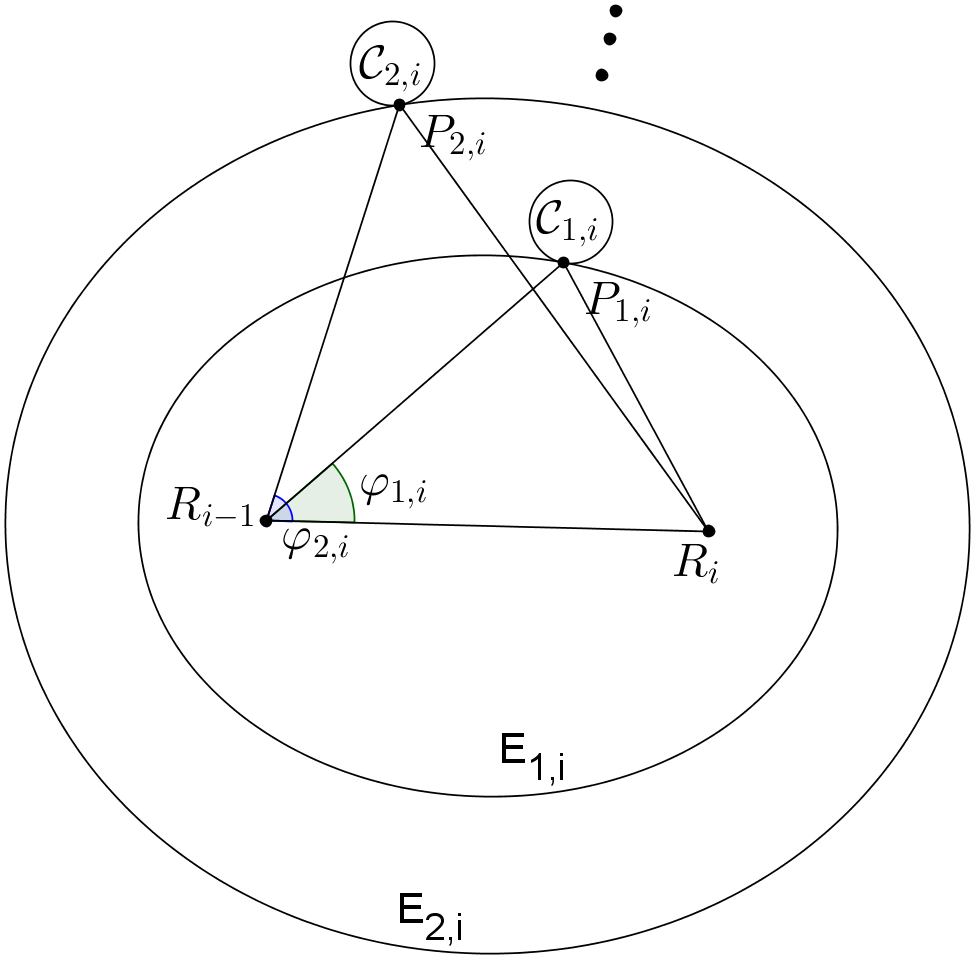} 
\includegraphics[clip,height=5.7cm]{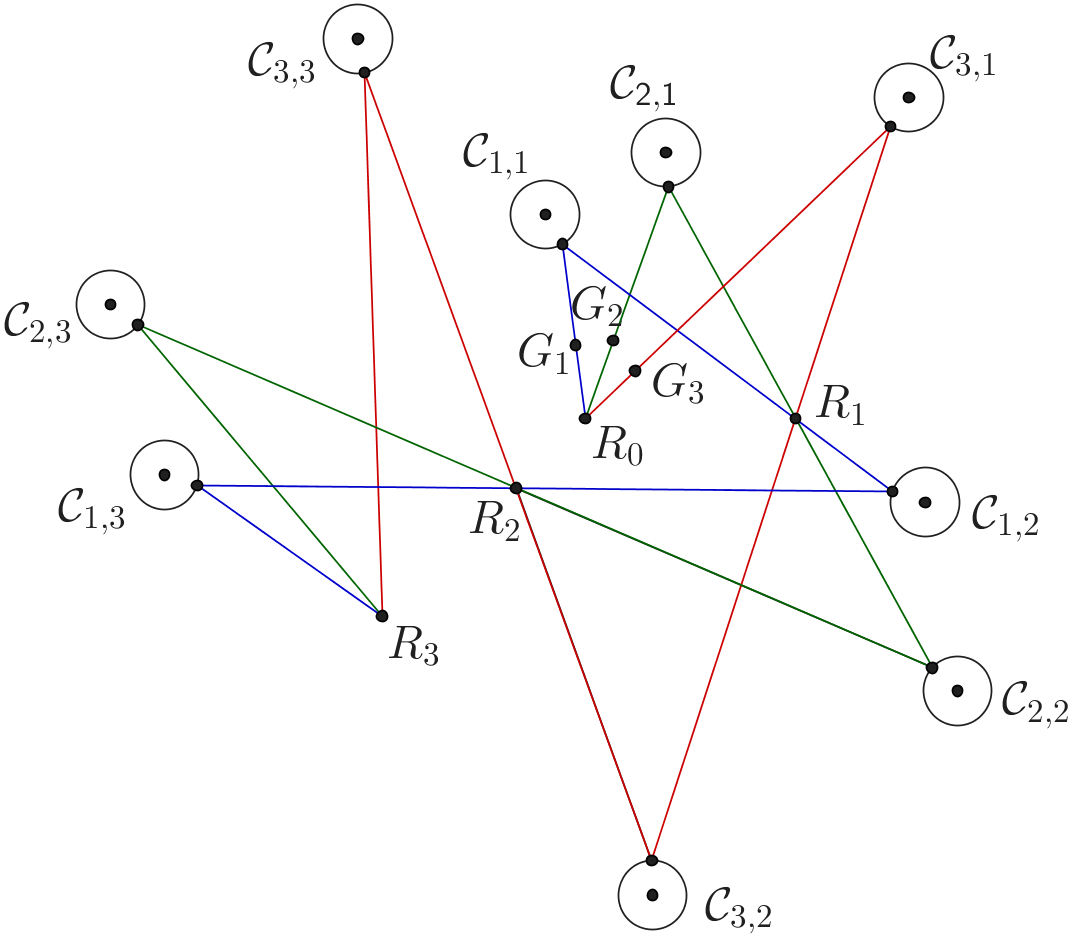} 
} 
\caption{ 
\textcolor{black}{\footnotesize{ 
On the left we see two receptors with the cones that redirect the light beams represented by continuous black lines. The cones are placed on the elipses $E_{k,i}$. On the right an example with $n=m=3$  (ellipses hidden). In the pictures, each circle $\mathcal{C}_{i,k}$ represents one of the cones that deviates the light beams and the $G_j$'s are possible spots one can place the grating grooves on. }}
}
\end{figure}

Now, if a straight line makes an angle $\varphi_{k,i}$ with the semi-major axis of the ellipse $E_{k,i}$, it intersects the ellipse at some point $P_{k,i}$. If we could just reflect the light ray at point $P_{k,i}$ and make it travel the extra distance $\epsilon_{k,i}$, we would have the light beam traveling the desired distance $b_{k,i}\lambda_{i}-b_{k,i-1}\lambda_{i-1}$ between $R_{i-1}$ and $R_{i}$. As will be shown in detail over the next section, this can be achieved by making the light ray surround a well situated cone $\mathcal{C}_{i,k}$. If one does that, then it is just a matter of choosing $\varphi_{k,1}$, $k=1,...,n$ and situating each $G_{k}$ in the segment between the source $O=R_{0}$ and $P_{k,1}$.

\subsection{Geometry of Cones}
\label{sub:GeometryCones}

Given a circular sector of radius $R$ and inner angle $\theta$, we may fold it into a cone of base radius $r$ and height $h$. The arclength associated to the sector is $R\theta$ radians, is equal to the perimeter $2\pi r$ of the base of the folded cone, hence
\begin{equation} \label{rpequeno}
r=\frac{R\theta}{2\pi}.
\end{equation}

Therefore, from the Pythagorean Theorem applied to the basis radius $r$, the generatrix $R$ and the height $h$ of the cone, we have

\begin{equation} 
 \label{hpequeno} h=R\sqrt{1-\left(\frac{\theta}{2\pi}\right)^{2}}. 
\end{equation}

Since the cone is locally flat, to describe the path of a light ray on its surface we may look at its unfolded image on the circular sector. Since the two (green) edges of the sector are glued together, a line that joins identified points on each edge will became a closed path on the cone. 
\begin{figure}[!ht]
\label{fig: circular sector and cone}
\center{
        \includegraphics[height=6cm]{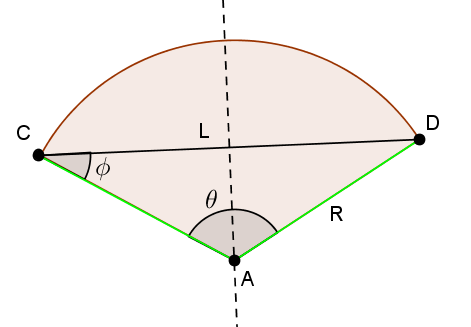}
        \includegraphics[height=6cm]{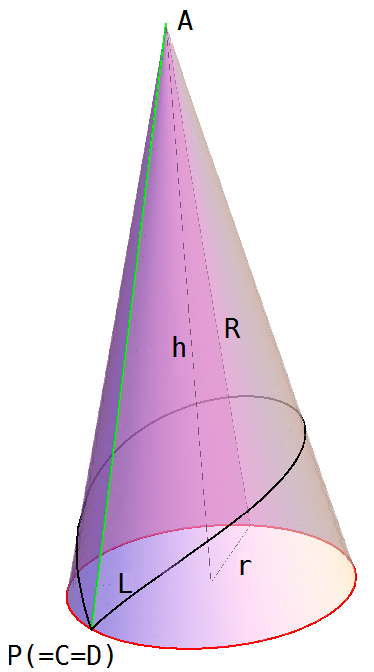}
}
        \caption{If one folds the circular sector on the left identifying
the two green lines the result is the folded cone on the right.}
\end{figure}

In particular, by symmetry along the bisecting line of the sector, a line coming through vertex $C$ and leaving through vertex $D$ is the unfolding of a geodesic on the cone that enters and leaves the cone at the same point $P$ (corresponding to $C=D$) with the same angle $\phi.$  Let us  denote  by $L:=\overline{CD}$ the length of this loop. 

By the Law of Sines:
\begin{eqnarray}
  \label{LporR}         \frac{L}{\sin\theta}         =         \frac{R}{\sin\left(\frac{\pi-\theta}{2}\right)}         &\Rightarrow&          L         =         2R\sin\left(\frac{\theta}{2}\right) 
\end{eqnarray}

Furthermore, since our light ray comes from the surrounding plane and climbs the cone, we must make sure that the ray entering the cone follows the prescribed geodesic. To do that, we must compute the angle $\phi$ between the side of the sector and the line $(CD)$, that is

\begin{equation} \label{angulo_phi}
 \phi=\frac{\pi-\theta}{2}
\end{equation}

\begin{figure}[!ht] \begin{center}\includegraphics[height=5cm]{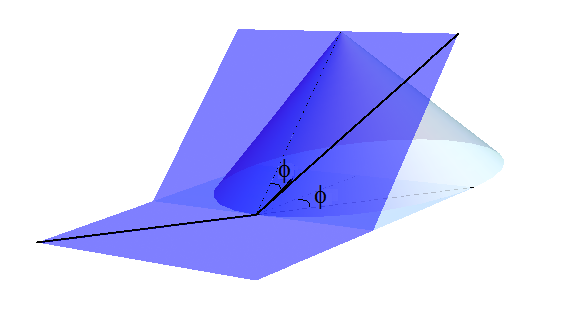} \end{center} \caption{\textcolor{black}{\footnotesize{The ray of light climbing a bent plane is the local picture at the point of contact between the flat and the conical section of the surface $\sigma_0$.}}} \label{fig: tangent plane} 
\end{figure}

Since the folding is an isometry, the angle between the geodesic and the generatrix at the incidence point $P$ is still $\phi$. Finally, taking the tangent plane of the cone at $P$ along the folding edge  (Figure \textcolor{black}{\ref{fig: tangent plane}})  reveals that the angle between the incident ray and the outer normal of the cone base measures also  $\phi$.

\subsection{The Cone $\mathcal{C}_{k,i}$} \label{TheCone}

Let us design the cone that fits our situation.  In order to have the light beam follow the sides of the triangle $\Delta_{k,i}:=R_{i-1}P_{k,i}R_{i}$, we put the center point of the cone base, $c_{k,i}$ on the bisector of the angle $2\phi_{k,i}=\widehat{R_{i-1}P_{k,i}R_{i}}$. We want a geodesic like the one in Figure \ref{fig: circular sector and cone} to follow a path of length $\epsilon_{k,i}$ along the cone. So form the cone from a circular sector of radius $\rho_{k,i}$ and angle $\theta_{k,i}$. Applying equations (\ref{LporR}) and (\ref{angulo_phi}) we have: 
\begin{equation}        \label{eq: rho and theta} 
\begin{cases}        
 \displaystyle         \rho_{k,i}=\frac{\epsilon_{k,i}}{2\cos\phi_{k,i}}\\         \theta_{k,i}=\pi-2\phi_{k,i}\\ \end{cases}
. 
\end{equation}
 Therefore the folded cone $\mathcal{C}_{k,i}$ has basis radius $r_{k,i}$ and height $h_{k,i}$ given by relations (\ref{rpequeno}) and (\ref{hpequeno}):
\[ \begin{cases}
 \displaystyle r_{k,i}=\frac{\epsilon_{k,i}(\pi-2\phi_{k,i})}{4\pi\cos\phi_{k,i}}\\ \displaystyle h_{k,i}=\frac{\epsilon_{k,i}}{2\cos\phi_{k,i}}\sqrt{1-\left(\frac{\pi-2\phi_{k,i}}{2\pi}\right)^{2}} \end{cases}.
 \]
We place $c_{k,i}$ on the bisector of $2\phi_{k,i}$ at a distance $r_{k,i}$ away from the triangle $\Delta_{k,i}$ (see section \ref{TheCone}). Simplifying for $\phi_{k,i}$ and denoting by $e_{k,i}$ the eccentricity of ellipse $E_{k,i}$, we find \begin{equation} 
       \label{eq: phi} \phi_{k,i}=\frac{1}{2}                 \left\{                 \pi-\varphi_{k,i}         -\arccos\left[                 \displaystyle                 \frac{(a_{k,i}^{2}+d_{i}^{2})e_{k,i}\cos\varphi_{k,i}-2d_{i}^{2}}{2e_{k,i}^{2}a_{k,i}^{2}\cos\varphi_{k,i}-(a_{k,i}^{2}+d_{i}^{2})e_{ki}}                 \right]                 \right\} 
\end{equation}
 with $a_{k,i},\ d_i$ and $\varphi_{k,i}$ as in section \ref{Elipse}. We have thus explicitly calculated all the parameters of an admissible array of the  $\{R_{i}\}$.

\subsection{Conformal mapping onto the plane disk}

\begin{figure}
[!ht]
 \begin{center}
         \includegraphics[width=\textwidth]{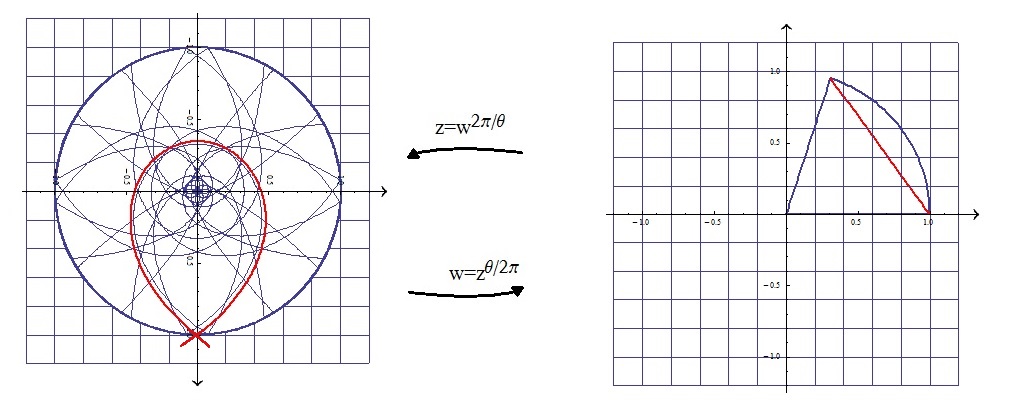} 
\end{center}
 \caption{\textcolor{black}{\footnotesize{The one-to-one mapping $z=w^{2\pi / \theta}$ takes conformally the euclidean circular sector that could be folded into a cone onto the  disk with a non-flat metric.}}} \label{conformeMapa} 
\end{figure}

In order to complete the procedure presented at section \ref{procedimento}, we must conformally map the disk on the plane that forms the cone's base onto the cone itself. Since the mapping of the disk sector onto the cone is an isometric parametrization, it suffices to map the disk of radius $r_{k,i}$ conformally onto the disk sector of radius $\rho_{k,i}$ and angle $\theta_{k,i}$.
If we settle the origin  at the center of the circle and identify the plane with $\mathbb{C}$, there is a well-known way of creating a conformal mapping of the form $w(z)=Kz^{\frac{1}{\alpha}}$ between a circle and a circular sector.  Given the sector, we first rescale it to the unit disk by  $ z_1(w)=\frac{w}{\rho_{k,i}}.$ Then, map it onto the open unit disk (up to a branching line) by $  z_2(w)=\displaystyle w^{\frac{2\pi}{\theta}}$ as shown in Figure \ref{conformeMapa}. Finally, we rescale the disc onto the original disk of radius $r_{k,i}$ using $ z_3(w)=r_{k,i}w $. The map of the sector onto the disk is given by
\begin{equation*}
z(w)=(z_3\circ z_2\circ z_1)(w)=r_{k,i}\left(\frac{w}{\rho_{k,i}}\right)^{\frac{2\pi}{\theta_{k,i}}}
\end{equation*}

In his famous paper on the theoretical framework for invisibility devices \cite{leonhardt_optical_2006}, Leonhardt shows that the ratio of refractive indices before ($\alpha$) and after $(\alpha')$ a conformal mapping $w(z)$ is given by  $$ \eta_w(z):=\frac{\alpha}{\alpha'}= \left\vert \frac{dw}{dz} \right\vert. $$ In our context, the inverse mapping is defined almost everywhere on the disk by
\begin{equation*}
w(z)=\rho_{k,i}\left(\frac{z}{r_{k,i}}\right)^{\frac{\theta_{k,i}}{2\pi}}
\end{equation*}
and, since $|z|\leq r_{k,i}$, we may substitute back equations (\ref{eq: rho and theta}) and (\ref{eq: phi}):
\begin{equation}        \label{eq: refractive index ratio}         
\eta(z)         =         \frac{\theta_{k,i}\rho_{k,i}}{2\pi}\left(\frac{|z|}{r_{k,i}}\right)^{\frac{\theta_{k,i}}{2\pi}-1}         \leq          \frac{\theta_{k,i}\rho_{k,i}}{2\pi}
\end{equation}

\section{Example: Modular Array}

To illustrate the solution and get some idea of the orders of magnitude involved in a real device, we propose a pet model: an array in which all the receptors $R_{i}$ lie on the horizontal axis with $d_{1}=\lambda_{1}$, $d_{i}=\lambda_{i}-\lambda_{i-1},\forall i\in{\{2,...,m\}}$. Also, we choose $b_{k,i}=k$, that is:
\[ \overline{OG_{k}}+\overline{G_{k}R_{i}}=k\lambda_{i},\forall i\in{\{1,...,m\}} \]
Note that we may also put the first groove point $G_{1}$  on the axis of receptors, so $\epsilon_{1,i}=0,\forall i$ and there is no cone $\mathcal{C}_{1,i}$. In order to have the same $\varphi_{k,i}$ for all $i$, allowing a modular setting, we take $\epsilon_{k,i}\propto\left(\lambda_{i}-\lambda_{i-1}\right)$ for $k>1$, say $\displaystyle\epsilon_{k,i}:=\frac{\lambda_{i}-\lambda_{i-1}}{n}$. Then, for $k>1$, we have:
\[ a_{k,i}=\frac{(kn-1)(\lambda_{i}-\lambda_{i-1})}{2n}. \]
By symmetry each $c_{k,i}$ will be in the bisector of $\overline{R_{i-1}R_{i}}$, so  $$ \varphi_{k,i}=\arccos\frac{n}{kn-1}  \quad\text{and}\quad \phi_{k,i}=\frac{\pi}{2}-\varphi_{k,i}. $$
The parameters of the cone $\mathcal{C}_{k,i}$ are:
\[ \begin{cases} \displaystyle  r_{k,i}=\frac{2(\lambda_{i}-\lambda_{i-1})\arccos\frac{n}{kn-1}}{4\pi n\sin\left(\arccos\frac{n}{kn-1}\right)}=\frac{(\lambda_{i}-\lambda_{i-1})\arccos\frac{n}{kn-1}}{2\pi n\sqrt{1-\left(\frac{n}{kn-1}\right)^{2}}}\\ \displaystyle h_{k,i}=\frac{\lambda_{i}-\lambda_{i-1}}{2n\sqrt{1-\left(\frac{n}{kn-1}\right)^{2}}}\sqrt{1-\left(\frac{\arccos\frac{n}{kn-1}}{2\pi}\right)^{2}}\\ \theta_{k,i}=2\varphi_{k,i} \end{cases} \] and its center is on the bisector of $\overline{R_{i-1}R_{i}}$ at a distance $b_{i,k}=\sqrt{a_{k,i}^{2}-\left(\frac{\lambda_{i}-\lambda_{i-1}}{2}\right)^{2}}$ from the line.

\begin{figure}[ht] \begin{center} \includegraphics[height=7cm]{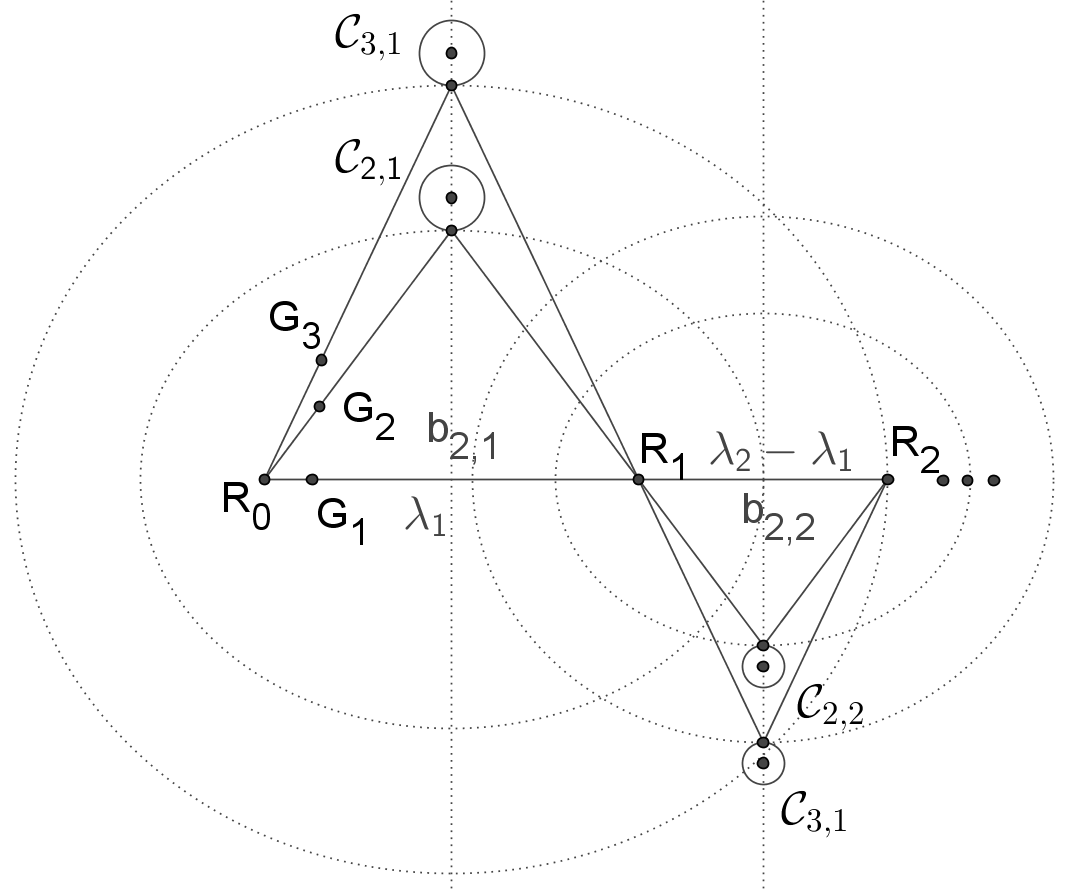} \end{center} \caption{The depiction of the module for the distribution of cones, receptors and possibly the samples in the modular array described. The straight black lines represent the light rays. The dotted ellipses are simply part of the construction steps.} \end{figure}

In this construction, we made a simulation with the following constraints and parameters: \begin{itemize}         \item          200 wavelengths ($\lambda_i$) equally spaced between $1.4\mu m$ and $1.6 \mu m$;         \item         Distance between two receptors ($R_j$) greater than $2\mu m$;         \item         Distance between grating grooves ($G_k$) greater than $2\mu m$; \end{itemize} Varying the distance between receptors, that is the most restraining parameter to the model, we tried to keep the refractive index within the order of magnitude of 1, but even calculating to the extreme case of $\epsilon_{k,i}=b_{k,i}\lambda_i-b_{k,i-1}\lambda_{i-1}$, i.e. collapsing the ellipses' foci, only letting unthinkable $2\, m$ as the distance between receptors, which would result in a $2\, km$ radius device to accomodate all 200 receptors, we could set the refractive index on the sub-optimal range between $0.2$ and $1.2$. From an engineering viewpoint, this is unfortunately quite far from actual device construction at the moment. Nevertheless, the mathematical solution presented in Section \ref{sec general solution} is general and may be improved for practical applications in the future. Also, a wider range of refractive indexes may be possible soon, as suggested in \cite{pendry_controlling_2006}.

\section*{Acknowledgements}
We thank Professor Ricardo Mosna for valuable ideas in the development of this project. HS is supported by Fapesp research grant 2014/24727-0 and by CNPq Productivity PQ2 grant 312390/2014-9. VS was supported by Fapesp grant 2012/21923-7.

\bibliography{bibliography_NOVO}

\end{document}